# Wake response to an ocean-feedback mechanism: Madeira Island case study

R.M.A. Caldeira[1,2,*] and R. Tomé[3,4]

[1] CCM-Center for Mathematical Sciences, University of Madeira, Campus da Penteada, 9000-390 Funchal, Madeira, Portugal

[2] CIIMAR-Interdisciplinary Centre of Marine and Environmental Research, Rua dos Bragas, 289, 4050 - 123 Porto, Portugal.

[3] CCMMG- Centro do Clima, Meteorologia e Mudanças Globais, University of Azores, Pólo Universitário de Angra do Heroísmo, 9701-851, Angra do Heroísmo, Portugal.

[4] Instituto Don Luis (IDL), Faculdade de Ciências da Universidade de Lisboa Campo Grande, Edifício C8, Piso 3, 1749-016, Lisboa, Portugal

*Correspondence to: rcaldeira@ciimar.up.pt

**Abstract**

This discussion focused on the numerical study of a wake episode. The Weather Research and Forecasting model was used in a downscale mode. The current literature focuses the discussion on the adiabatic dynamics of atmospheric wakes. Changes in mountain height and consequently on its relation to the atmospheric inversion layer should explain the shift in wake regimes: from a 'strong-wake' to a 'weak-wake' scenario. Nevertheless, changes in SST variability can also induce similar regime shifts. Increase in evaporation, contributes to increase convection and thus to an uplift of the stratified atmospheric layer, above the critical height, with subsequent internal gravity wave activity.

## 1. Introduction

Madeira Island is located in the NE Atlantic (32.5ºN; 17ºW), with a volcanic origin its mountain range dominates the island's interior orography. The mountains are oriented perpendicular to the incoming trade winds. Therefore, the interaction between the mountains with the predominant NE flows, combined with the fact that the atmospheric inversion layer is often located bellow the mountain top (1500-1800m), nurtures the formation of atmospheric wakes, frequently manifested as Von Karman Vortex Streets (VKVS, hereafter). In fact, images of Madeira VKVS have inspired many studies in the classic scientific literature (e.g. Chopra and Hubert, 1965; Chopra, 1973; Scorer, 1986; Etling, 1989; Etling, 1990).

In contrast with neighboring archipelagos (e.g. Canaries), the Madeira Island atmospheric wake is dominated by the interaction of one island, since the nearby islands (Desertas and Porto Santo), lay well bellow the atmospheric inversion layer (200-300m) (see Fig. 1). As previously discussed in Etling (1989), the wake becomes a two-dimensional flow problem if the 'dividing streamline height' ($h_c$), remains bellow the mountain-top (H), warrantying that is the inversion height ($Z_i$) remains bellow the ($h_c$); whereas it becomes three-dimensional one, if the 'dividing streamline height' passes over the maintain-top. The 'dividing stream line height' was defined as:

$$hc = H(1 - Fr) \quad (1)$$

Where the Froude Number (Fr) is represented as:

$$Fr = \frac{U}{NH} \quad (2)$$

Smith et al (1997), taking into account previous works (Schar and Smith, 1993; and Grubisic et al., 1995), refined Etling (1989), concepts, re-classifying wake scenarios into four main types (i) steady wake eddies; (ii) eddy-shedding; (iii) long-straight wake and (iv) no-wake. In Smith et al (1997) proposed concepts, the 'diving stream line height' ($h_c$), became the critical height for mountain wave breaking. In brief, the new definition was based on the amount of Potential Vorticity (PV) generated by the perturbed (island) flow. PV was generated only if the mountain height (H) was greater than ($h_c$); for slightly greater values weak-wakes were formed, whereas for considerably greater differences strong PV generation would justify eddy generation and self-advection i.e. strong-wakes. Nevertheless, since these concepts were based on shallow-water model theory, leeward eddies could vary from stationary (Re< $Re_c$) to periodic shedding cases (Re >$Re_c$). Critical Reynolds ($Re_{c)}$ values were determined in laboratory conditions to be approximately 50, considering the islands as cylindrical obstacles.

Different archipelagos were used to illustrate the diversity of wake regimes, Hawaii was proposed to be a good representation of steady wake case, the Aleutians were proposed to be a periodic eddy shedding example; St. Vincent was presented as long-straight wake case and the Barbados was referred as the no-wake case. Nevertheless, meteorological conditions change over time and strong and weak wake episodes, may co-occur on the same island.

Miranda et al., (1999), used a combination of microbarographs and numerical models to study daily pressure oscillations in the northern and southern coasts of Madeira Island. These differences were attributed to gravity wave drag effects. Diabatic drag effects due to the coupling between tropospheric greenhouse gases and the surface was briefly discussed as an alternative reasoning, but unsubstantiated by the lack of observational evidences. Prior to that Miranda and Valente (1997), had studied the critical resonance in flows passing isolated Gaussian Mountains, using a three-dimensional linearized model, and associated the strong internal atmospheric wave breaking to the critical-level dynamics.

Most recently however, Small et al., (2008), in a review of air-sea interaction over ocean fronts and eddies, suggested that the ocean forces the atmosphere through heat fluxes out of the ocean. Differences in SST induces differences in humidity concentration, this in turn leads to changes in near-surface stability and surface stress as well as latent and sensible heat fluxes (Sweet et al., 1981; Businger and Shaw, 1984; Hayes et al., 1989). A change in stability modifies the logarithmic profiles within the sea surface layer, with consequential responses to the vertical gradients of wind velocity, air-potential temperature and relative humidity (Small et al., 2008). This SST feedback mechanism can lead to changes in the transfer of momentum and consequently to changes in the atmospheric boundary layer height. Nevertheless, most studies have emphasized the atmospheric response to SST over small-scale gradients i.e. eddies and fronts. Xie (2004) emphasized the different atmospheric over cool waters from the response over warm waters, suggesting that at larger scales, episodes like El-Nino can modulate deep-convection.

In order to study the variations of wake regimes in the Madeira Region, the first effort was to replicate a realistic episode (control experiment). The second study focused on the change of mountain maximum height and consequently on the change of the inversion layer height in relation to the mountain-top (aka critical height). This alluded to a change in the overall flow regime and allowed the representation of strong and weak-wake scenarios as previously proposed by Smith et al (1997) and Etling, (1989). To introduce thermodynamic variability, a third study focused on the wake response to changes in sea surface temperature. Discussion focused on the feedback effect of SST during wake formation.

## 2. Numerical model configuration

In the present study, the Advanced Research Weather and Forecasting Model, version 3.2, was setup using three one-way nested domains. All domains were centered in Madeira Latitude using Lambert projection (32.5ºN; 16.5ºW). The top of the model was located at 50 hPa and a total of 60 vertical sigma-levels were considered. All grids used NOAH Land Surface Model, the RRTMG Scheme for long and shortwave radiation, the Dudhia Shortwave and the Rapid Radiation Transfer Model (RRTM). Longwave Radiation and the WSM 6-class graupel scheme were used for the cloud physics. The Mellor-Yamada-Janjic TKE scheme was used for the Planetary Boundary Layer and the Betts-Miller-Janjic was employed in the 54 and 18km grids. The initial and time dependent boundary conditions were derived from the ECMWF ERA-Interim reanalysis, for the study period (3-10 of July, 2002). Reanalysis products are produced at 6-hour time intervals, with 0.5 degrees spatial resolution, and 22 vertical pressure levels (previously interpolated at Mars ECMWF Data Server). The three model domains are represented in figure 2.

Five numerical experiments were conducted: (i) the control experiment used the ERA-Interim reanalysis product and a sigma-interpolated version of island realistic orography. The changes in the island orography were obtained by multiplying the original dataset by a constant factor; (ii) in the high-mountain experiment, the orographic data was multiplied by a factor o 1.5, in order to increase mountain height, further above the inversion layer; and (iii) by a factor of 0.5, in order to lower all the island-mountains bellow the inversion layer, i.e. low mountain experiment. To study the thermodynamic impact on wake formation, the realistic representation of Madeira orography was maintained (identical to the control experiment), and the SST field was modified in WRF. For the (iv) warm case, (+5ºK) was added to the WRF calculated mean SST field; whereas for the (v) cold case, (-5ºK) was subtracted from the mean WRF calculated SST field. The objective of the later exercise was to study the impact of the changes in SST, in inducing changes in the wake scenarios.

## 3. Results and discussion

### 3.1 Parameter space

Table 1 compares the main parameters, which distinguishes the different numerical experiments. Measures of $U_0$ and N were extracted from a model grid-point windward of the island (34.2257ºN; 17.0219 º W). Wave breaking will occur for a greater (non-dimensional) reference height of:

$$\hat{h} = \frac{hN}{U_0} \tag{3}$$

The parameter ($\hat{h}$) is also a measure of the amplitude of nonlinear effects by the mountain, often referred to as the inverse Froude number. For sufficient high values of ($\hat{h}$), mountain waves will break; the critical value for two-dimensional flow is $\hat{h} \sim 0.85$. As can bee seen from Table-1, $\hat{h}$ is significantly greater than unity for the warm case i.e. 5.399 thus suggesting strong internal gravity wave activity.

Another important parameter to consider is the Richardson number ($Ri$):

$$Ri = \frac{N^2 D^2}{4 U_0^2} \tag{4}$$

Where D represents the depth of the shear layer. The Richardson number expresses the ratio of potential to kinetic energy; in fluid dynamics, Ri values much greater than unity suggests that vertical mixing is dominant and therefore there is insufficient kinetic energy available to homogenize the fluid. Thus strong vertical mixing is expected to occur in the warm study case where Ri~7.3.

Table 1 – Calculated parameters for the studied cases

|  | Control | High | Low | Cold | Warm |
|---|---|---|---|---|---|
| Umax [m s$^{-1}$] | 25.1 | 25.1 | 25.1 | 25.3 | 25.7 |
| Hmax [m] | 1025.9 | 1538.8 | 512.9 | 1025.9 | 1025.9 |
| Zi [m] | 872 | 872 | 872 | 482 | 929 |
| N [s$^{-1}$] | 0.006 | 0.018 | 0.006 | 0.023 | 0.173 |
| $\hat{h}$ | 0.120 | 0.368 | 0.123 | 0.483 | ==5.399== |
| Ri | 0.004 | 0.034 | 0.004 | 0.058 | ==7.287== |

### 3.2 Momentum mediated wake

Results show an intensification of the surface vorticity (strong wake), generated by the tallest Madeira Island. In strong wakes, potential vorticity generated by the mountain advects itself as eddies, whereas in the weak-wake case it is simply advected downstream (figure 3). Weak-wake conditions were also studied in the lee of St. Vicent and other Windward Islands of the southeastern Caribbean (Smith et al., 1997). The noteworthy aspect of weak-wakes, apart from the absence of reverse flow, is its remarkably straight look (figure 3). Similarly, in weak conditions the Madeira Wake, it extends itself further leeward, showing no eddy shedding. Smith et al (1997) also suggests a leeward reacceleration of the flow, caused by the ambient pressure gradient rather than by lateral entrainment of geostrophic flows, as it occurs in the strong-wake cases.

### 3.3 Diabatic mediated waves

Weak wake scenarios can therefore be developed under warm SST conditions. Figure 4 shows vertical profiles of relative humidity, in the transect leeward (south) of Madeira, comparing the warm, cold and control experiments. Northern/windward transect (not shown) were less perturbed than the leeward/southern one. Relative humidity was used as a proxy for atmospheric stratification and vertical mixing. As can be observed in figure 4, the atmospheric stratified layer is less perturbed in the control and cold experiments, and highly perturbed in the warm case. The water vapor contribution is far greater in warm SST conditions favoring convective cell activity, which in turn contributed to the erosion of atmospheric stratification. The plots also show that the vertical perturbations of the first layer are discontinuous, suggesting some wave activity in the surrounding region. Figure 5 shows an iso-surface representation of the relative humidity, in the cold and warm experiments. As predicted by Smith et al (1997), V-shape and leeward waves are formed in the weak-wake case (warm case), such as the ones described for St. Vincent Island. The model also replicates well the VKVS (strong wake), case for the controlled experiment, without any apparent wave activity, in the surrounding region. There might be a resonant effect of the convective cells formation in the warm case, which might also contribute to the formation of atmospheric leeward waves.

### 3.4 Resonant waves

Comparing the cross-section of potential temperature in a (north-south), transect over the island, with a parallel transect over the open-ocean, two types of perturbed flows can be distinguished. Whereas in the over-the island transect weak wake scenarios show a strong over-the-mountain flows as originally proposed by Etling (1989) and by Smith (1997), including the formation of spurious internal gravity wave activity. The cross-section over the ocean revealed that the warm wake was strongly amplifying the wave resonant effect (Figure 6). Resonant waves have an amplitude of ~11Km and a period of 2.5 hrs. Overlaying this high-frequency variability, there are also waves with daily and weekly periods, in all studied cases.

It is hypothesized that the near-field energy damping is greater in the warm case than it is in the other cases, since the transfer of energy between the two systems (potential and kinetic), is done at very different frequencies than the wake-induced frequencies. This was somewhat suggested by the very different values assumed by (Ri) in the warm case, also indicating a strong discrepancy between the two storage modes.

### 3.5 Oceanic feedback mechanisms in wake flows

The cloud clearing affect of island wakes surrounded by dense clouds in open ocean-ocean regions is commonly observed around Madeira. Consequently, an oceanic warm wake is formed (see Caldeira et al., 2002). Figure 7a represents the SST monthly mean (July 2002), showing a warm SST signature, leeward of Madeira Island. Sequential atmospheric wake episodes of heating and cooling of the sea surface might feedback onto the atmosphere contributing to the observed change of wake scenarios. Increase in surface evaporation, contributes with high contents of water vapor to the lower atmospheric boundary layer, thus stimulating convective mixing and therefore contributing to the weakening the original atmospheric stratification developed during the strong wake episodes. This warm wake effect seems to persist during the summer months as depicted by the warm SST signature depicted by the seasonal mean (figure 7b).

Koseki et al., (2008), studied global diabatic and adiabatic effects of orographically forced climate. Results based on changing mountain heights and SST values on a numerical simulation showed that, the sensible heat response to mountain uplift is similar to that induced by evaporation i.e. convective flows.

The study also suggests that stationary waves forced by orography, drive oceanic scale gyres, which in turn strengthens the diabatic forcing of the atmosphere, acting to amplify the stationary waves, with visible impacts on the surrounding regions. Prior to that several studies had also suggested the link between thermally generation of convective flows and the generation of atmospheric internal gravity waves (e.g. Lintner and Neelin, 2007; Aves and Johnson, 2008).

In a study relating SST with wind speed over mesoscales, Small et al (2005), concluded that wind speed response varied from 0.5 to 1.5 ms$^{-1}$ per ºC of SST change. Considering that the leeward wake of Madeira produces a cumulative SST difference between 1.5 and 2 ºC, wind speeds are expected to respond to these SST changes, particularly near the island flanks.

In the study of St. Vincent Island, Smith et al., (1997) considered that the warming of the surface boundary might have caused a downward turbulent mixing of potentially warmer air from aloft. It was also proposed that the hydraulic jump induced by wave braking was the main source of such vertical mixing. Nevertheless, the diabatic explanation never gained substantial support due to the lack of observational evidences. Herein we propose diabatic flow effects in the context of the Madeira wake phenomena. Due to sea surface heating, the strong Madeira wake episodes are subsequently weakened by strong convective flows, uplifting the stratified layer until a weak wake and/or dissipation (no-wake) state is achieved.

## 4. Conclusions and future work

In view of the current analysis a new conceptual model is proposed for the Madeira atmospheric wake dynamics (figure 8). Strong wake episodes are formed. In strong stratified atmospheric condition the flow becomes bi-dimensional, forcing the wind to contour the mountains. Nevertheless, in time these strong flows contribute vortex shedding, due to wake formation, moving patches of stratocumulus leeward, and in a cyclic manner, exposing the leeward side of the island to stronger incident solar radiation, comparatively to the surrounding open-ocean regions. Warmer sea surface temperatures increase the local evaporation rates contributing to the increase vertical mixing in the lower atmosphere. Vertical mixing in turn, contributes to the erosion of the stratified first layer, enabling the incoming wind to start to flow over the mountain, thus changing the wake regime from strong to weak wake. Initial periods of change from strong to weak wake conditions are expected to be very energetic, manifested by the formation of resonant internal gravity waves. Future studies should combine in situ, satellite and numerical model results to investigate the time and spatial evolution of Madeira induced atmospheric wakes. Attempts should be made to consider consecutive wake regimes and not solely dominant scenarios. The study of the dynamics of island wakes could provide important insights in the air-sea interaction which occurs at the oceanic (sub)mesoscales.

**Acknowledgements**
This work was carried out in the scope of two research projects funded by the Portuguese National Science Foundation (POCI/MAR/57265/2004 and PPCDT/MAR/57265/2004). We would also like to thank Annick Terpstra and Gert-Jan Steeneveld, from Wageningen Universiteit, Netherlands, for the inspiring report. We acknowledge the MODIS mission scientists and associated NASA personnel for the production of the SST data used herein.

**References**
Aves S.L., and Johnson R.H. 2008. The diurnal cycle of convection over the northen South China Sea. *Journal of the Meteorological Society of Japan* 86, 6:919-934.

Businger, J.A. Shaw W.J. 1984. The response of the marine boundary layer to mesoscale variations in sea-surface temperature. *Dyn. Atmos. Oceans* 8: 267–281.

Caldeira R.M.A. Groom S. Miller P. Pilgrim D Nezlin N. 2002. Sea-surface signatures of the island mass effect phenomena around madeira island, northeast atlantic. *Remote Sensing of Environment* 80: 336 – 360.

Chopra K.P. and Hubert L.F. 1965. Mesoscale Eddies in Wake of Islands. *Journal of the Atmospheric Sciences* 22: 652-757.

Chopra K.P. 1973. Atmospheric and oceanic flow, problems introduced by islands. *Advances in Geophysics* 16: 297-421.

Elting D. 1989. On Atmospheric Vortex Streets in the Wake of Large Islands. *Meteorol. Atmos. Phys.* 41: 157-164.

Elting, D. 1990. Mesoscale Vortex Shedding from Large Islands: A Comparison with Laboratory Experiments of Rotating Stratified Flows. *Meteorol. Atmos. Phys.* 43: 145-151.

Grubišić V. Smith R. B. and C. Schär. 1995. The effect of bottom friction on shallow-water flow past an isolated obstacle. *J. Atmos. Sci.* 52:1985–2005.

Hayes S.P. McPhaden M.J. Wallace J.M. 1989. The influence of sea surface temperature on surface wind in the eastern equatorial. Pacific: weekly to monthly variability. *J. Climate* 2: 1500–1506.

Koseki S. Watanabe M. Kimoto M. 2008. Role of the Midlatitude Air-Sea interaction in orrographically forced climate. *Journal of the Meteorological Society of Japan* 86:335-351.

Lintner, B.R. and Neelin J.D. 2007. Time scales and spatial patterns of passive ocean-atmosphere decay modes. *Journal of Climate* 21: 2187-2203.

Miranda P.M.A. and Valente M.A. 1997. On Critical Level Resonance in Three-dimensional Flow Past Isolated Mountains. *Journal of the Atmospheric Sciences* 54, 12:1574-1588.

Miranda P.M.A. Ferreira J.J. and Thorpe A.J. 1999. Gravity wave drag produced by Madeira. *Quarterly Journal of the Royal Meteorological Society* 125: 1341-1357.

Scorer R., 1986. *Cloud Investigation by Satellite*. Chichester: Ellis Horwood, 300pp.

Schär C. and Smith R. B. 1993. Shallow-water flow past isolated topography. Part I: Vorticity production and wake formation. *J. Atmos. Sci.* **50:** 1373–1400.

Scorer, R. S. 1986. Cloud investigation by satellite, chap. Cloud investigation by satellite, John Wiley, New York.

Smith R.B. Gleason A. Gluhosky P. and Grubišić V. 1997. The wake of St. Vincent. *J. Atmos. Sci.,* **54,** 606–623.

Small R.J. Xie S. P. Hafner J. 2005. Satellite observations of mesoscale ocean features and copropagating atmospheric surface fields in the tropical belt. *J. Geophys. Res.* 110: C02021.

Small R.J. deSzoeke S.P. Xie S.P. O'Neill L. Seo H. Song Q. Cornillon P. Spall M. Minobe S. 2008. Air–sea interaction over ocean fronts and eddies. *Dynamics of Atmospheres and Oceans*. 45, 3–4: 274-319.

Sweet W. Fett R. Kerling, J. La Violette, P. 1981. Air–sea interaction effects in the lower troposphere across the north wall of the Gulf Stream. *Mon. Wea. Rev.* 109: 1042–1052.

Xie, S.P. 2004. Satellite observations of cool ocean–atmosphere interaction. *Bull. Am. Meteor. Soc*. 85: 195–208.

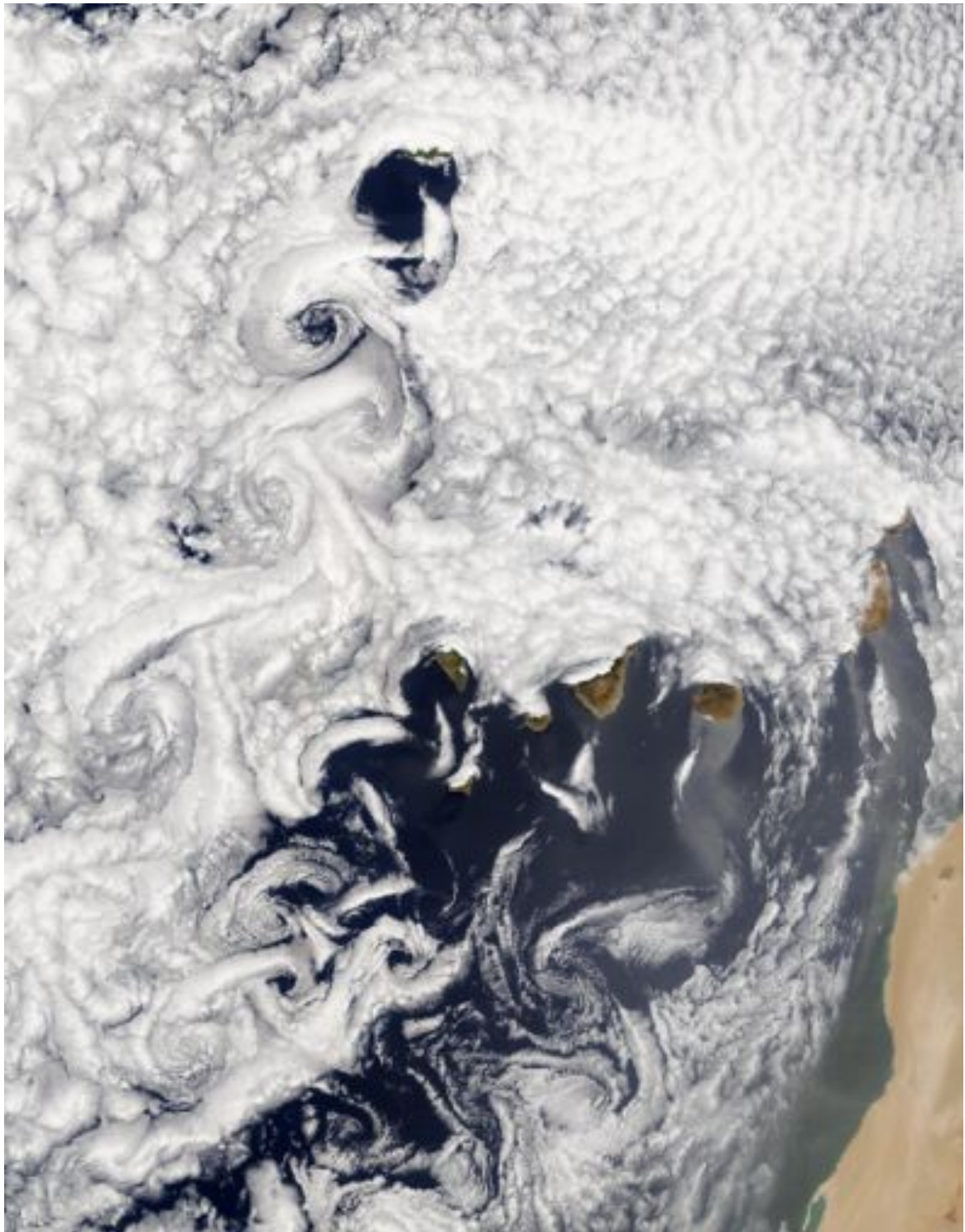

Figure 1 – MODIS Aqua satellite image (visible), capturing an intense VKVS forming leeward of Madeira Island (north) and which went further south than the Canaries Archipelago. Image captured on the 5th of July 2002; 11:55 UTC.

(a)

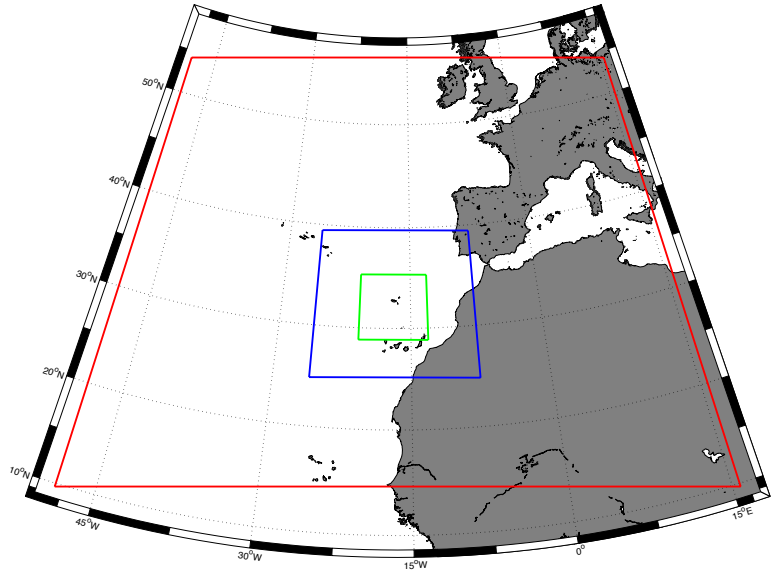

(b)

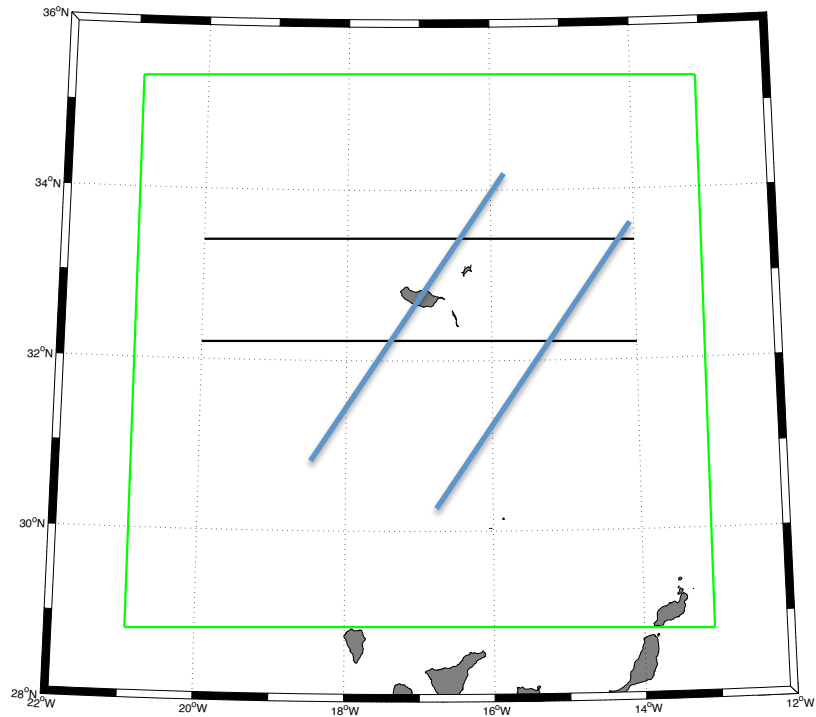

Figure 2 – WRF calculated domains. (a) Three domains were considered: (a) 54km (red); 18km (blue) and 6km (green); (b) Higher resolution domain representing upstream and downstream cross-section locations (black lines – figure 4). Blue lines represent north-south transects used to represent the cross-sections in figure 6.

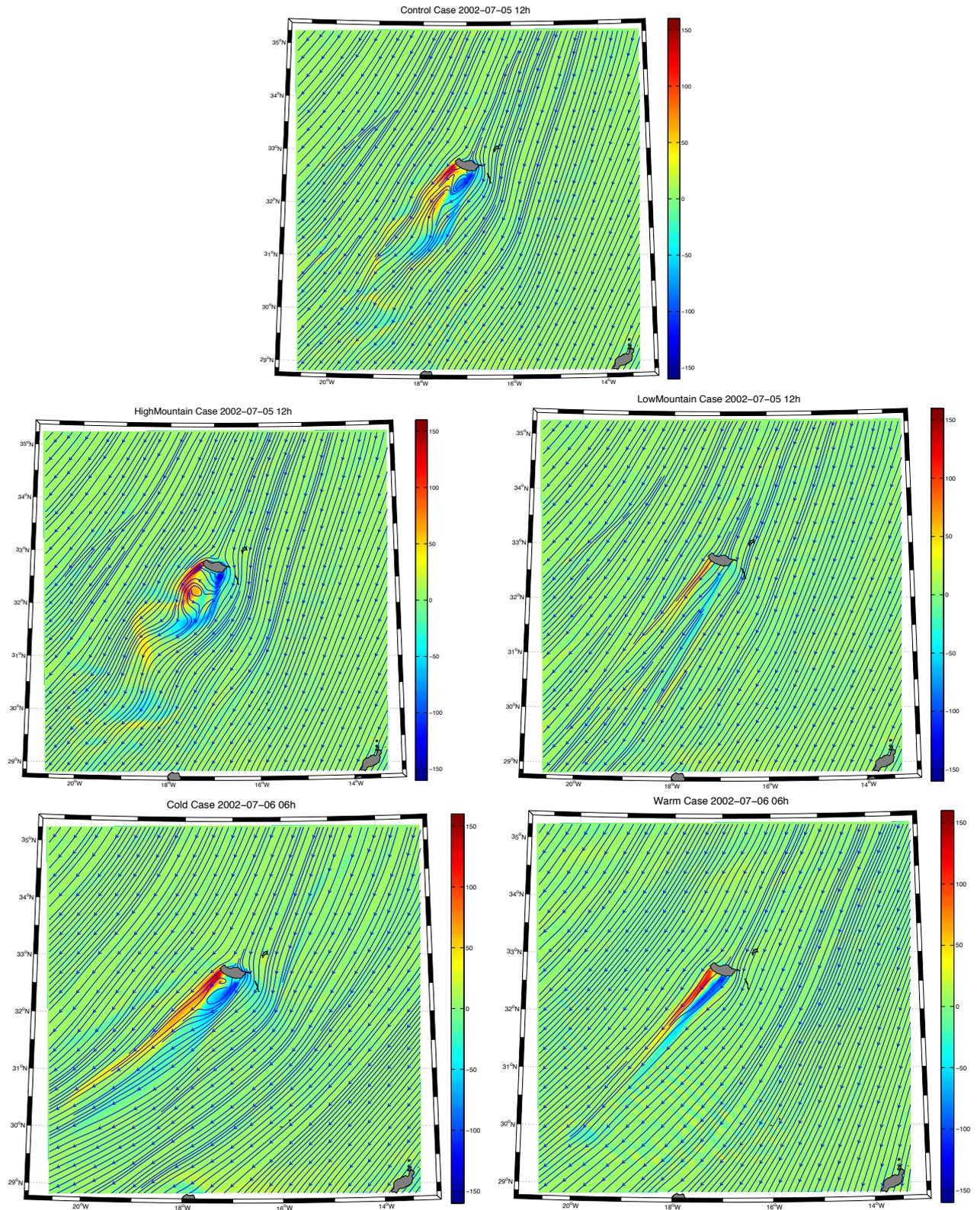

Figure 3 – Relative vorticity and streamlines representative of the five case studies:
(a) Control; (b) High-mountain; (c) Low mountain; (d) Cold and (e) Warm cases.

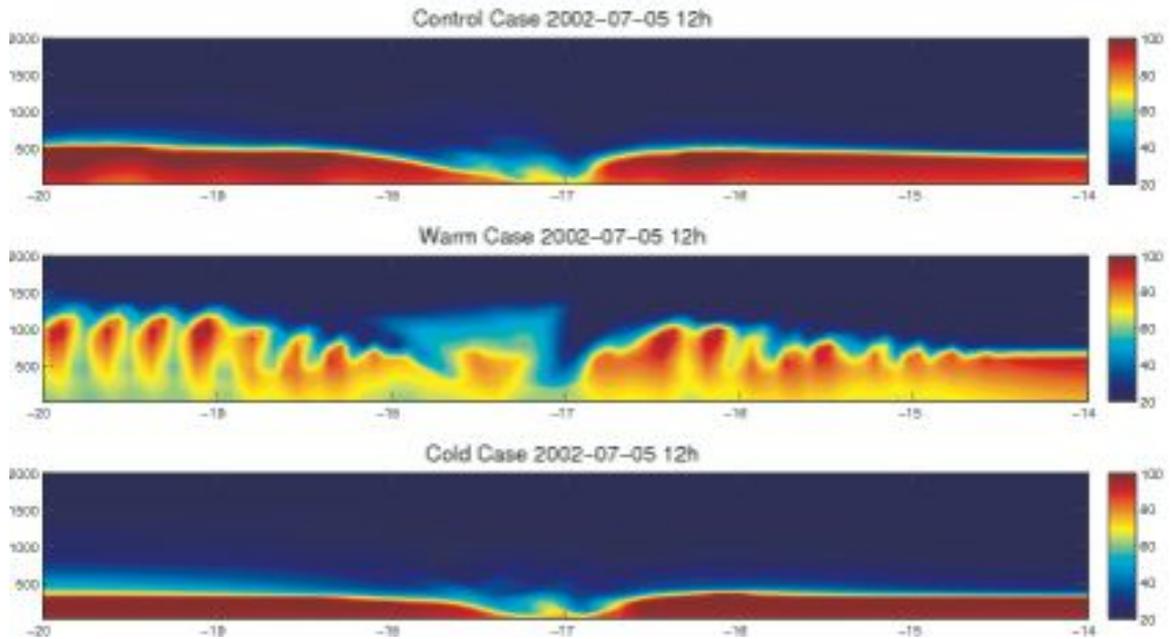

Figure 4a – Cross-sections of relative humidity, in the Southern transect (figure 2), leeward of the island for the (a) control; (b) warm; and (c) cold cases.

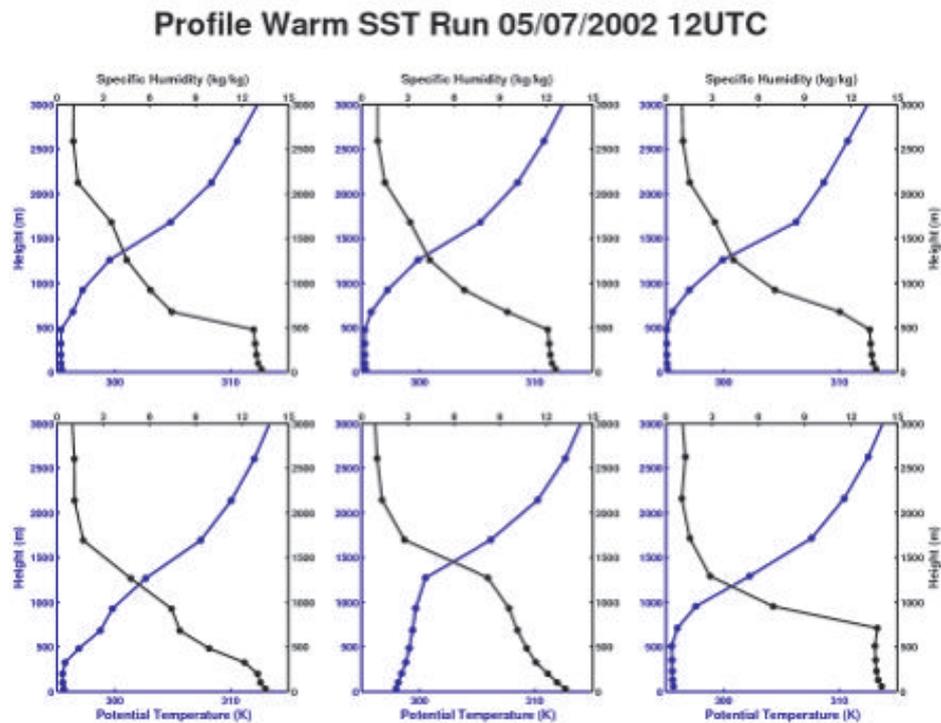

Figure 4b – Vertical profiles of specific humidity and potential temperature; for windward (top panels) and leeward (bottom panels), of Madeira Islands. Profile locations correspond to the center and Western and Eastern extremes of the cross-sections transect shown in figure 2.

(a)
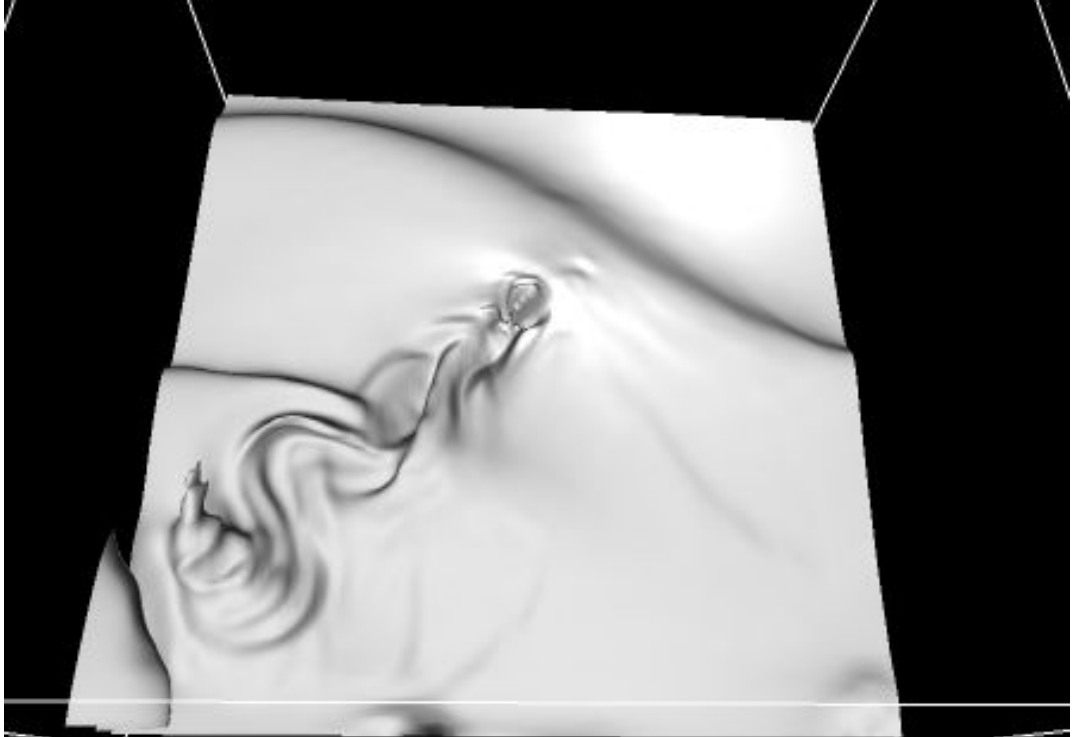
(b)
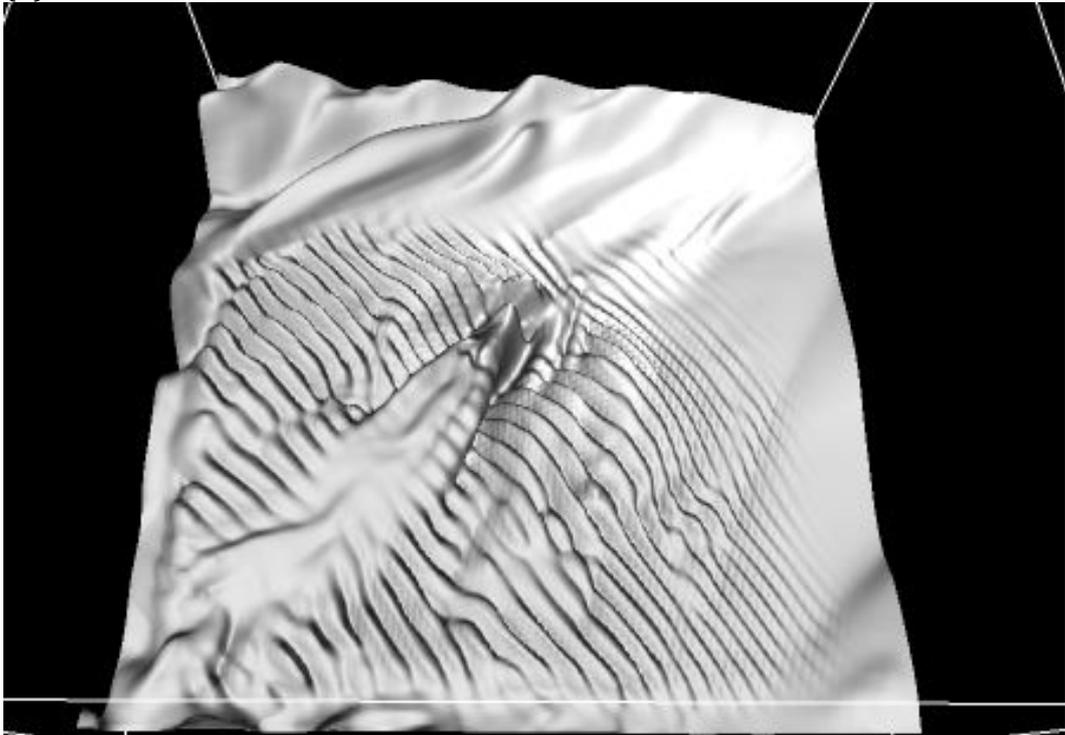

Figure 5 – Volume integrated plot of relative humidity (over 30 sigma layers), leeward of the island comparing the (a) cold; and (b) warm case studies.

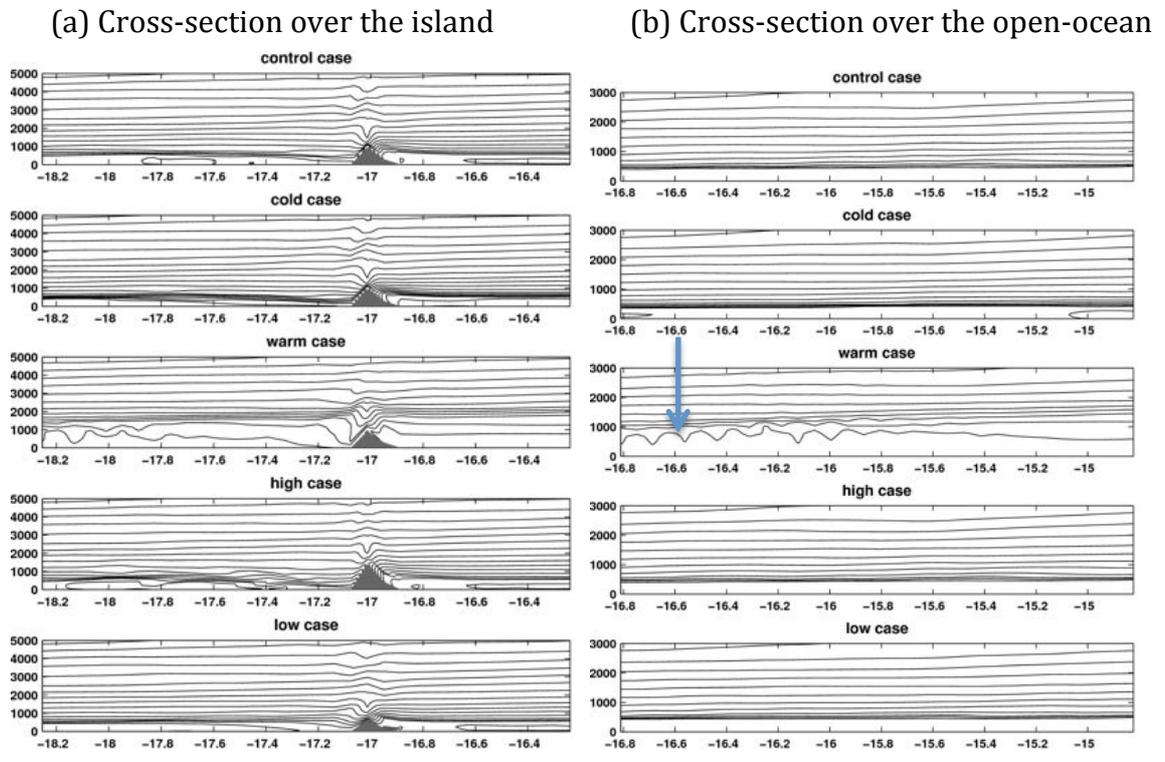

Figure 6 – Cross sections (height in meters) of potential temperature for the 6th of July 2002 showing (a) strong over-the-island perturbed flows for the weak-wake cases (warm and low); (b) strong resonant wave activity (blue arrow) present in the warm wake case, over the surrounding open-ocean as depicted by figure 5.

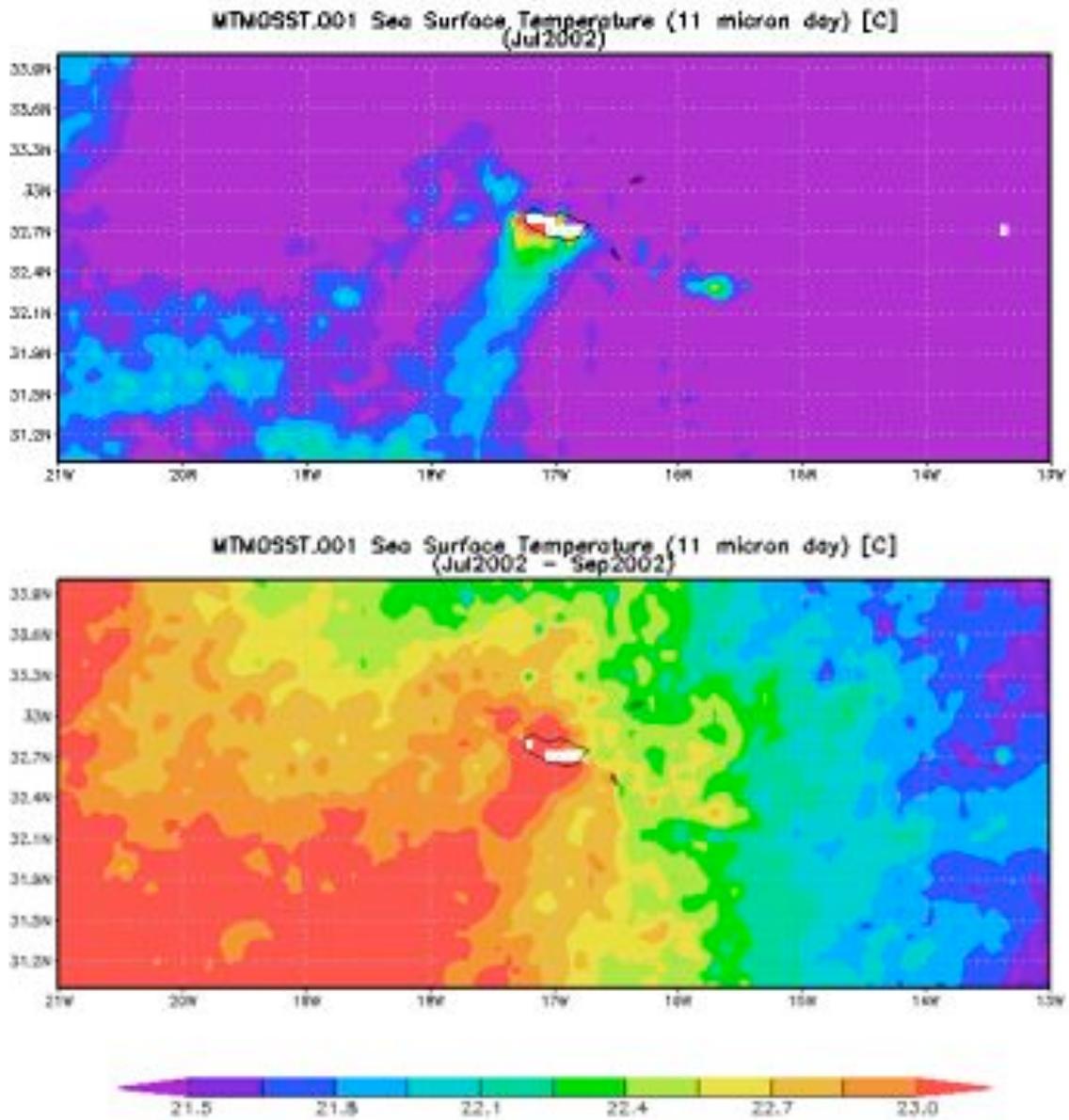

Figure 7 – Mean SST computed for (a) July 2002; and for (b) summer 2002 (July-September) using MODIS-Aqua, satellite level-3 data at 9km spatial resolution. Leeward warm water differentiates the leeward of Madeira in 2ºC relatively to the surrounding oceanic region, indicating the cumulative effect of successive wake events.

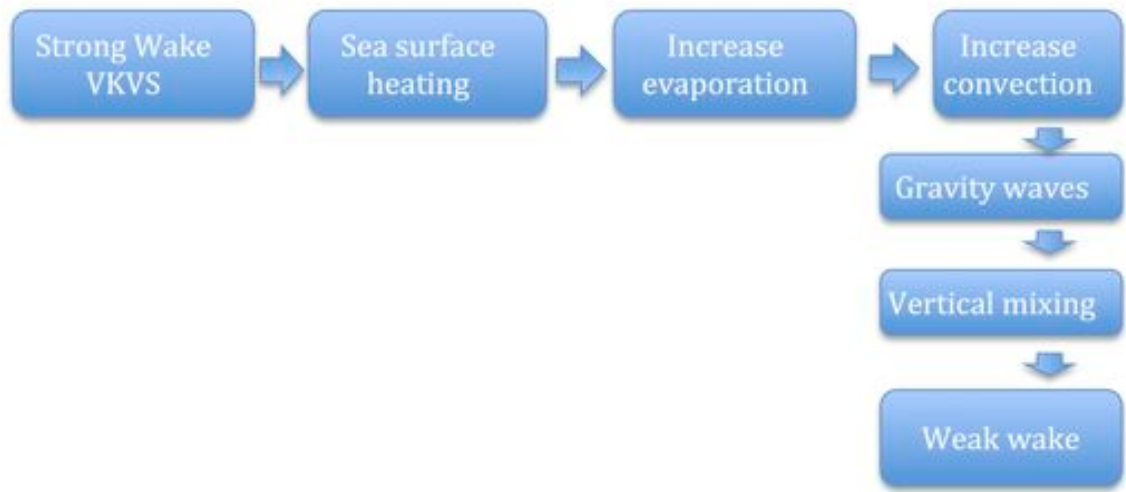

Figure 8 – New proposed conceptual model for the Madeira Atmospheric wake dynamical evolution, in response to sea surface feedback mechanism